\documentclass[aps,prd,preprint,superscriptaddress]{revtex4}

\usepackage{graphicx}
\usepackage{amsmath,amssymb}
\usepackage{bm}
\usepackage{color}

\def\ee{\end{eqnarray}}

\def\=:{=\hspace{-.7em}\raisebox{1.1ex}{.}\hspace{.1em}\raisebox{-0.2ex}{.} }

\def\ee{\end{eqnarray}}

\def\=:{=\hspace{-.7em}\raisebox{1.1ex}{.}\hspace{.1em}\raisebox{-0.2ex}{.} }

\newcommand {\beq}{\begin{eqnarray}}
\newcommand {\eeq}{\end{eqnarray}}
\newcommand {\non}{\nonumber\\}

\newcommand {\1}[1]{\frac{1}{#1}}

\newcommand {\ph}{\varphi}

\newcommand {\del}{\partial}

\newcommand {\tr}{{\rm tr}\,}

\begin{document}


\title{Non-Abelian Sine-Gordon Solitons
}


\author{Muneto Nitta}

\affiliation{
Department of Physics, and Research and Education Center for Natural 
Sciences, Keio University, Hiyoshi 4-1-1, Yokohama, Kanagawa 223-8521, Japan\\
}


\date{\today}
\begin{abstract}
We point out that non-Abelian sine-Gordon solitons stably exist 
in the $U(N)$ chiral Lagrangian. 
They also exist in 
a $U(N)$ gauge theory 
with two $N$ by $N$ complex scalar fields 
coupled to each other. 
One non-Abelian sine-Gordon soliton can terminate on 
one non-Abelian global vortex.  
They are relevant in chiral Lagrangian of QCD or
in color-flavor locked phase of  high density QCD,
where the anomaly is suppressed at asymptotically high temperature 
or density, respectively.

\end{abstract}

\maketitle

\section{Introduction}
Sine-Gordon kinks (solitons) \cite{Perring:1962vs}
appear in broad range of physics 
from classical and quantum field theories 
\cite{Manton:2004tk,Rajaraman:1982is},
QCD \cite{Eto:2013hoa}, 
conformal field theories, 
integrable systems, 
and cosmology \cite{Vilenkin:2000}
to condensed matter physics. 
Condensed matter systems offer a lot of examples 
of sine-Gordon kinks 
which can be observed in laboratory experiments, 
such as
Josephson junctions of two superconductors \cite{Ustinov:1998},
those in multi-layer high $T_{\rm c}$ superconductors \cite{Blatter:1994},
two-gap superconductors 
\cite{Tanaka:2001,Gurevich:2003,Goryo:2007}, 
chiral p-wave superconductors \cite{Agterberg1998}, 
coherently coupled two-component Bose-Einstein condensates (BECs) 
\cite{Son:2001td}, 
 two separated BECs with a Josephson coupling \cite{Kaurov:2005},
helium 3 superfluids \cite{Volovik2003}, 
and ferromagnets \cite{Chen:1977}.
In particular, sine-Gordon kinks are 
Josephson vortices 
in Josephson junctions appearing 
when a magnetic field is applied parallel to 
a Josephson junction or 
layers of  high $T_{\rm c}$ superconductors 
\cite{Ustinov:1998,Blatter:1994,Kaurov:2005,Nitta:2012xq}.
Another interesting case is that 
a sine-Gordon kink connects two fractional vortices 
winding around different components, 
to constitute a vortex molecule 
in multi-gap superconductors
\cite{Babaev:2001hv,Goryo:2007,Nitta:2010yf} 
and coherently coupled multi-component BECs
\cite{Kasamatsu:2004,Cipriani:2013nya,Eto:2012rc}.

Sine-Gordon kinks also explain relations between 
topological defects or solitons in different dimensions. 
Since-Gordon kinks inside the world-volume of 
a topological defect 
represent some other topological defects in the bulk; 
Sine-Gordon kinks inside a domain wall 
are vortices, lumps or baby Skyrmions in the bulk 
\cite{Auzzi:2006ju,Nitta:2012xq,Kobayashi:2013ju,Jennings:2013aea},
which explains a relation between 
sine-Gordon kinks and ${\mathbb C}P^1$ instantons
\cite{Sutcliffe:1992ep,Stratopoulos:1992hq}.
Sine-Gordon kinks inside a domain wall ring 
are baby Skyrmions \cite{Kobayashi:2013ju}. 
They represent Skyrmions in the bulk if residing in a domain wall 
within a domain wall 
\cite{Nitta:2012wi,Gudnason:2014nba,Nitta:2012rq}
or in a vortex string \cite{Gudnason:2014hsa,Gudnason:2014gla},  
they are Hopfions in the bulk if residing in a toroidal domain wall \cite{Kobayashi:2013xoa}, and
are Yang-Mills instantons in the bulk if residing inside a monopole string 
in Yang-Mills theory in $d=4+1$ dimensions 
\cite{Nitta:2013cn}.

There have been many proposal of generalizations 
of the sine-Gordon model.
One of such is a complex sine-Gordon model 
describing a vortex motion in superfluids 
\cite{Lund:1976ze},
the $O(4)$ model \cite{Pohlmeyer:1975nb}, 
conformal field theories
\cite{Bakas:1993xh}, and 
a domain wall junction \cite{Naganuma:2001br}.
There have been non-Abelian generalizations such as
the matrix sine-Gordon model 
\cite{Park:1995rj},
the symmetric space  sine-Gordon model 
\cite{Bakas:1995bm} 
and so on.

In this paper, we discuss yet another non-Abelian generalization of 
sine-Gordon kinks.
We point out that 
a non-Abelian sine-Gordon kink admitted in 
the $U(N)$ chiral Lagrangian 
\cite{Gepner:1984au} is a non-Abelian soliton 
carrying non-Abelian moduli  
${\mathbb C}P^{N-1} \simeq SU(N)/[SU(N-1)\times U(1)]$.
Here, the term ``non-Abelian" is used in the same way with 
that of non-Abelian vortices \cite{Hanany:2003hp,Auzzi:2003fs,
Shifman:2004dr,Eto:2005yh} 
carrying non-Abelian ${\mathbb C}P^{N-1}$ moduli,
see Refs.~\cite{Tong:2005un,Eto:2006pg,Shifman:2007ce} for 
a review.
As in the same manner with 
a non-Abelian vortex with non-Abelian moduli  
which can terminate on a non-Abelian monopole 
because of the matching of the moduli ${\mathbb C}P^{N-1}$  
\cite{Auzzi:2003em,Eto:2006dx},
 non-Abelian sine-Gordon kink here can terminate 
on a non-Abelian global vortex 
\cite{Balachandran:2002je,Nitta:2007dp,Nakano:2007dq,Eto:2009wu},
see Ref.~\cite{Eto:2013hoa} as a review.
We then promote the non-Abelian sine-Gordon solitons to 
those in non-Abelian $U(N)$ gauge theories 
with two $N$ by $N$ complex scalar fields 
coupled to each other by a non-Abelian extension of 
linear or quadratic Josephson interaction. 
The Abelian case reduces to  
phase solitons in two-gap superconductors 
\cite{Tanaka:2001,Gurevich:2003,Goryo:2007}, 
while the non-Abelian extension is relevant to 
a color superconductor of the color-flavor locking phase 
of dense QCD matter
\cite{Alford:2001dt,Eto:2013hoa}.

This paper is organized as follows.
In Sec.~\ref{sec:model}, 
after reviewing sine-Gordon kinks
 in the conventional sine-Gordon model,
we discuss non-Abelian sine-Gordon kinks
in the $U(N)$ chiral Lagrangian.
In Sec.~\ref{sec:model2}, 
sine-Gordon kinks with a modified mass term 
and their non-Abelian $U(N)$ generalization 
are discussed.
In Sec.~\ref{sec:gauge}, 
these sine-Gordon kinks are promoted to 
gauge theories. 
The $U(1)$ gauge theory is nothing but 
two-gap superconductors 
or chiral p-wave superconductors 
corresponding to the conventional or 
modified mass term, respectively.
In Sec.~\ref{sec:NA-vortex}, we discuss that 
a sine-Gordon kink can terminate on a 
non-Abelian global vortex. 
Sec.~\ref{sec:summary} is devoted to 
summary and discussion.

\section{The sine-Gordon model and chiral Lagrangian \label{sec:model}}

\subsection{The sine-Gordon model}

The sine-Gordon kink is characterized by
the first homotopy group $\pi_1[U(1)] \simeq {\mathbb Z}$.
The Lagrangian density of conventional sine-Gordon model is 
\beq
 {\cal L} &=& \1{2} (\del_{\mu} \theta)^2 
- m^2 \left(1-\cos \theta\right) 
 \label{eq:SG}
\eeq
with $\mu=0,1,\cdots,d-1$ and $0 \leq \theta <2\pi$.
We consider static configurations depending on one spatial direction $x$.
The static energy density is 
\beq
 {\cal E} &=& \1{2} (\del_x \theta)^2 
+ m^2 \left(1-\cos \theta\right) .
\eeq
The Bogomol'nyi completion for the energy density  is obtained as 
\beq
 {\cal E} 
&=& \1{2} (\del_x \theta)^2 + 2 m^2 \sin^2{\theta \over 2} \non
 &=& \1{2} \left(\del_x\theta \mp 2 m \sin {\theta\over 2}\right)^2 
  \pm 2 m \partial_x \theta \sin {\theta\over 2} \non
 &\geq& 
 \left|2 m \partial_x \theta \sin {\theta\over 2}  \right|
 = \left|t_{\rm SG}\right|
\eeq
with the topological charge density defined by
\beq
t_{\rm SG} 
 \equiv 2m \partial_x \theta \sin {\theta\over 2}
 = - 4 m \partial_x \left( \cos {\theta \over 2}\right).
\eeq
The inequality is saturated by the BPS equation
\beq
 \del_x \theta \mp 2 m \sin {\theta\over 2} = 0.
\eeq
A single-kink solution interpolating between 
$\theta =0$ at $x \to -\infty$ to 
$\theta =2\pi$ at $x \to +\infty$ 
can be given as
\beq
 \theta(x) = 4 \arctan \exp{m (x- X)} 
\eeq
with the position $X$ in the $x$-coordinate.  
The topological charge for this solution is 
\beq
 T_{\rm SG} = \int dx t_{\rm SG} 
= -4 m \left[\cos {\theta\over 2}\right]^{x=+\infty}_{x=-\infty}
= -4 m (-1-1) = 8 m. 
\label{eq:SG-charge}
\eeq
The width of the sine-Gordon kink is $1/m$.

For later convenience, we introduce a new variable taking a value 
in the $U(1)$ group by
\beq
u \equiv e^{i\theta}.
\eeq
From $\del_x \theta = -(i/2) (u^* \del_x u - (\del_x u^*) u )$, 
the BPS equation is rewritten as
\beq
 -{i \over 2} (u^* \del_x u - (\del_x u^*) u )
 \mp m \sqrt {2-u-u^*} = 0 
\label{eq:BPS-U(1)}
\eeq
and the topological charge density is rewritten as
\beq
 t_{\rm U(1)} 
&=& -{i m\over 2} (u^* \del_x u - (\del_x u^*) u )
       \sqrt {2-u-u^*}
= - 2 m \del_x  \left( \sqrt {2+ u+u^*} \right) 
\eeq
The single-kink solution is
\beq
 u(x) = \exp \left(4 i \, \arctan \exp [m  (x- X)] \right) 
\label{eq:U(1)-one-kink}
\eeq
with the boundary condition $u \to 1$ for $x \to \pm \infty$.

\subsection{Non-Abelian sine-Gordon model 
as chiral Lagrangian}
Here we consider the $U(N)$ group:
\beq
U(x) \in U(N) \simeq {U(1) \times SU(N) \over {\mathbb Z}_N} 
\eeq
with the first homotopy group is nontrivial:
\beq
 \pi_1 [U(N)] = {\mathbb Z}.
\eeq
The Lagrangian for a $U(N)$ principal chiral model (chiral Lagrangian) 
for a $U(N)$-valued field $U(x)$ is given by
\beq
{\cal L} &=& \1{2} \tr \del_{\mu} U^\dagger \del^{\mu} U - {m^2 \over 2} \tr (2{\bf 1}_N - U - U^\dagger) \non
&=& \1{2} \tr (i U^\dagger \del_{\mu} U)^2 - {m^2 \over 2}\tr (2{\bf 1}_N - U - U^\dagger). \label{eq:U(N)SG}
\eeq
This Lagrangian is invariant under the chiral 
$SU(N)_{\rm L} \times SU(N)_{\rm R}$ symmetry 
\beq
 U(x) \to V_{\rm L}U(x)V_{\rm R}^\dagger , 
\quad V_{\rm L,R} \in SU(N)_{\rm L,R} \label{eq:U(N)sym0}
\eeq
The Lagrangian admits the unique vacuum $U={\bf 1}_N$.
The chiral symmetry is spontaneously broken to 
the vector-like symmetry
\beq
 U(x) \to V U(x)V^\dagger , 
\quad V \in SU(N)_{\rm L+R=V} .\label{eq:U(N)sym}
\eeq

The energy density for static configuration and 
its Bogomol'nyi completion are given as 
\beq
{\cal E}
&=& \1{2} \tr (i U^\dagger \del_x U)^2 
     - {m^2\over 2} \tr (2{\bf 1}_N - U - U^\dagger)\non
&=& \1{2} \tr \left[- {i\over 2} (U^\dagger \del_x U -  \del_x U^\dagger U ) \mp  m \sqrt {2 {\bf 1}_N - U -U^\dagger} \right]^2\non
&& 
\pm {m \over 2} \tr \left[ - {i\over 2} (U^\dagger \del_x U -  \del_x U^\dagger U ) \sqrt {2 {\bf 1}_N - U - U^\dagger)}\right] \non
&\geq& |t_{U(N)}| ,
\eeq
with the topological charge, defined by
\beq
t_{U(N)} \equiv
- {m \over 2} \tr \left[ i (U^\dagger \del_x U -  \del_x U^\dagger U ) \sqrt {2{\bf 1}_N - U- U^\dagger}\right].
\eeq
The BPS equation is obtained as
\beq
 - {i\over 2} (U^\dagger \del_x U -  \del_x U^\dagger U ) \mp  m \sqrt {2 {\bf 1}_N - U - U^\dagger}
={\bf 0}_N. \label{eq:BPS-U(N)}
\eeq
This equation is invariant under the $SU(N)$ symmetry in Eq.~(\ref{eq:U(N)sym}). 

Let us construct solutions to this equation.
The simplest ansatz is given by the 
following {\it Abelian} solution
\beq
 U(x) = u(x) {\bf 1}_N . \label{eq:Abelian}
\eeq
By substituting this ansatz into  
 Eq.~(\ref{eq:BPS-U(N)}), we find that 
$u(x)$ again satisfies Eq.~(\ref{eq:BPS-U(1)}).
The tension (energy per unit area) 
of this configuration is $T=N T_{\rm SG}$.

Next, we construct {\it non-Abelian} solutions.
Let us consider the following ansatz \cite{Gepner:1984au}:
\beq
 U(x) = {\rm diag} (u(x),1,\cdots,1) \label{eq:ansatz}
\eeq
By substituting this ansatz 
into Eq.~(\ref{eq:BPS-U(N)}), we find that 
$u(x)$ satisfies Eq.~(\ref{eq:BPS-U(1)}) 
and the one-kink solution is obtained as Eq.~(\ref{eq:U(1)-one-kink}).
The tension of this configuration is $T=T_{SG}$. 
Although the solution is obtained by embedding the Abelian solution
into the upper-left corner, 
this solution is truly non-Abelian; 
In terms of group elements,
the ansatz in Eq.~(\ref{eq:ansatz}) can be rewritten as
\beq
 U(x) &=& \exp \left(i {\theta(x) \over N}\right) 
 \exp \left(i \theta(x) T_0 \right),\label{eq:T_0}\\
T_0 &\equiv& {1 \over N} {\rm diag.}(N-1,-1,\cdots,-1).
 \nonumber
\eeq
From this expression,
one can see that the $U(1)$ group element rotates only $2\pi/N$ 
while the rest is compensated by an $SU(N)$ group element $T_0$. 
Namely at $x=\infty$ ($\theta = 2\pi$) 
the $U(1)$ group element becomes
$\exp \left(i {2\pi \over N}\right) = \omega$ 
while the $SU(N)$ group element becomes 
$\exp \left( 2\pi i T_0 \right)
={\rm diag.} (\omega^{N-1},\omega^{-1},\cdots,\omega^{-1}) = \omega^{-1} {\bf 1}_N$.
The $SU(N)$ group element connects the trivial element to 
an element of the center ${\mathbb Z}_N$ of the $SU(N)$ group.

There is a continuous degeneracy of the solutions with the same energy. 
Since the Lagrangian and the BPS equation is invariant under 
the $SU(N)$ transformation in Eq.~(\ref{eq:U(N)sym}),
the most general solution is obtained as
\beq
 U(x) = V{\rm diag} (u(x),1,\cdots,1) V^\dagger , 
  \quad V\in SU(N).  \label{eq:SU(N)moduli}
\eeq
Since there exists a redundancy for the action of $V$, 
$V$ in fact takes a value in the coset space
\beq
 V \in {SU(N) \over SU(N-1) \times U(1)} \simeq {\mathbb C}P^{N-1}.
\eeq
Therefore, the one-kink solution has the moduli 
\beq
{\cal M} = {\mathbb R} \times {\mathbb C}P^{N-1}.
\eeq
In terms of the group elements, the general solution can be rewritten as
\beq
 U(x)
&=& \exp \left(i {\theta(x) \over N}\right) 
       \exp \left(i \theta(x)  VT_0 V^\dagger\right) \non
&=&  \exp \left(i {\theta(x) \over N}\right) \exp i{\theta(x)\over N}T,
\label{eq:general} 
\eeq
with $T \equiv V T_0 V^\dagger$.
$T$ can be any $SU(N)$ generator normalized as $e^{i2\pi T} = \omega^{-1}{\bf 1}_N$.

Let us introduce the orientational vector 
$\phi \in {\mathbb C}^N$ with a constraint
\beq
 \phi^\dagger \phi =1,
\eeq
which represents homogeneous coordinates of 
${\mathbb C}P^{N-1}$. 
The generator $T$ and the general solution in 
Eq.~(\ref{eq:general}) can be rewritten by using the orientational vector as 
\beq
&& T = V T_0 V^\dagger = \phi \phi^\dagger  - {1\over N}{\bf 1}_N, \\
&&  U(x)  = \exp \left({i\theta(x) \phi\phi^\dagger}\right).
\eeq

\section{The modified sine-Gordon model and chiral Lagrangian \label{sec:model2}}

\subsection{The modified sine-Gordon model}

We consider the Lagrangian density of a sine-Gordon model 
with an unconventional potential, given by 
\beq
 {\cal L} &=& \1{2} (\del_{\mu} \theta)^2 
- m^2 \left(1-\cos^2 \theta\right) 
 \label{eq:SG2}
\eeq
with $\mu=0,1$. 
This model admits two vacua $\theta=0,\pi$ 
in the defined range $0\leq \theta \leq 2\pi$.
We concentrate on static configurations.
The static energy density is 
\beq
 {\cal E} &=& \1{2} (\del_x \theta)^2 
+ m^2 \left(1-\cos^2 \theta\right) 
= \1{2} (\del_x \theta)^2 
+ m^2 \sin^2 \theta.
\eeq
The Bogomol'nyi completion for the energy density  is obtained as 
\beq
 {\cal E} 
&=& \1{2} \left[(\del_x \theta)^2 + 2m^2 \sin^2 \theta \right]  \non
 &=& \1{2} \left(\del_x\theta \mp \sqrt 2 m \sin \theta\right)^2 
  \pm \sqrt 2 m \partial_x \theta \sin \theta \non
 &\geq& 
 \left|\sqrt 2 m \partial_x \theta \sin \theta  \right|
= \left|t_{\rm SG}\right|
\eeq
with the topological charge density
\beq
t_{\rm SG} 
 \equiv \sqrt 2m \partial_x \theta \sin \theta
 = -\sqrt 2 m \partial_x \left( \cos \theta \right).
\eeq
The inequality is saturated by the BPS equation
\beq
 \del_x \theta \mp \sqrt 2m \sin \theta = 0.
\eeq
A one-kink solution interpolating between 
$\theta =0$ at $x \to -\infty$ to 
$\theta =\pi$ at $x \to +\infty$ 
can be given as
\beq
 \theta(x) = 2 \arctan \exp{\sqrt 2 m  (x- X)} 
\eeq
with the position $X$ in the $x$-coordinate  
and the width $1/m$.
The topological charge for this solution is 
\beq
 T_{\rm SG} = \int dx t_{\rm SG} 
= - \sqrt 2 m \left[\cos \theta \right]^{x=+\infty}_{x=-\infty}
= - \sqrt 2 m (-1-1) = 2 \sqrt 2 m. 
\label{eq:SG-charge2}
\eeq

In terms of $u(x)=e^{i\theta(x)}$, 
the BPS equation is rewritten as
\beq
&&\del_x u \mp {\sqrt 2  \over 2}m (1-u^2) = 0, \non
&& (\leftrightarrow  -i (u^* \del_x u - (\del_x u^*) u ) \mp \sqrt 2 i m (u-u^*) = 0  ),
\label{eq:BPS-U(1)2}
\eeq
and the topological charge density is rewritten as
\beq
 t_{\rm U(1)} &=& -  {\sqrt 2 \over 2}  m u^* \del_x u  (u-u^*)
\eeq
The one-kink solution is
\beq
 u(x) = \exp \left(2 i \, \arctan \exp{{\sqrt 2 m \over 4} (x- X)} \right).
\label{eq:U(1)-one-kink2}
\eeq

\subsection{Non-Abelian sine-Gordon model as 
chiral Lagrangian with modified mass}

The Lagrangian for $U(N)$ principal chiral model with a 
modified mass is
\beq
{\cal L} &=& \1{2} \tr \del_{\mu} U^\dagger \del^{\mu} U -V 
=  \1{2} \tr (i U^\dagger \del_{\mu} U)^2  - V \\
V 
&=& m^2 \tr(2{\bf 1}_N - U^2 - U^{\dagger2})
= m^2 \tr (2{\bf 1}_N - U - U^\dagger)(2{\bf 1}_N + U + U^\dagger)\non
&=& m^2 \tr ({\bf 1}_N - U^2) ({\bf 1}_N  - U^{\dagger2}).
\eeq
This model admits two vacua $U= \pm {\bf 1}_N$.
The energy density for static configuration and 
its Bogomol'nyi completion are given as 
\beq
{\cal E}
&=& \1{2} \tr \del_x U^\dagger \del_x U
     + m^2 \tr ({\bf 1}_N - U^2) ({\bf 1}_N  - U^{\dagger2})\non
&=& \1{2} \tr 
\left[ \{\del_x U^\dagger \mp \sqrt 2 m ({\bf 1}_N - U^{\dagger2})\} 
        \{\del_x U \mp \sqrt 2 m ({\bf 1}_N - U^2)\} 
\right] \non
&& \pm 2m \tr \left[  \del_x U^\dagger ({\bf 1}_N -U^2) + + \del_x U ({\bf 1}_N -U^{\dagger2}) \right] \non
&\geq& |t_{U(N)}|,
\eeq
with the topological charge, defined by
\beq
t_{U(N)} &\equiv&
\sqrt 2 m \tr \left[  \del_x U^\dagger ({\bf 1}_N -U^2) 
+ \del_x U ({\bf 1}_N -U^{\dagger2}) \right] \non
&=& \sqrt 2 m \tr \left[
U^\dagger \del_x U (U -U^{\dagger})  + {\rm h.c.}\right].
\eeq
The BPS equation is obtained as
\beq
&& \del_x U  \mp \sqrt 2 m ({\bf 1}_N -U^2) 
={\bf 0}_N \non
\leftrightarrow 
&&
(i U^\dagger \del_x U  \mp \sqrt 2 i m (U^\dagger -U) 
={\bf 0}_N ).
\label{eq:BPS-U(N)2}
\eeq

As in the same manner,
 the Abelian kink in Eq.~(\ref{eq:U(1)-one-kink2}) 
can be embedded into a conner as in Eq.~(\ref{eq:ansatz}) 
to obtain a non-Abelian kink.
Also, it allows the ${\mathbb C}P^{N-1}$ moduli as 
Eq.~(\ref{eq:SU(N)moduli}).

\section{Non-Abelian Sine-Gordon Soliton in Gauge Theories}\label{sec:gauge}

\subsection{Abelian gauge theory: 
two-gap superconductors and chiral p-wave superconductors}
Let us consider a $U(1)$ gauge theory 
coupled with two complex scalar fields $\phi_i(x)$ ($i=1,2$), given by
\beq
 {\cal L} = \1{2} \sum_{i=1,2} D_{\mu}\phi_i^* D^{\mu}\phi_i
 + {\cal L}_{\rm J}
 - \sum_{i=1,2} {\lambda_i \over 4}  (|\phi_i|^2 -1)^2
 + \1{4 e^2} F_{\mu\nu}^2 \label{eq:2compU(1)}
\eeq
with $D_{\mu} \phi_i = (\del_{\mu} - i A_{\mu} )\phi_i$.
${\cal L}_{\rm J}$ is a Josephson term either linear 
or quadratic: 
\beq
&&  {\cal L}_{\rm J,1} = {\gamma \over 2} (\phi_1^* \phi_2 + {\rm c.c.}-2) \non
&&  {\cal L}_{\rm J,2} = {\gamma \over 2} [(\phi_1^* \phi_2)^2 + {\rm c.c.}-2].
\label{eq:Josephson}
\eeq
The gauge transformation is defined by
\beq
&& \phi_i \to e^{i\alpha(x)}\phi_i, \quad 
 A_{\mu} \to A_{\mu} + \del_{\mu} \alpha(x), 
\eeq
while a $U(1)$ global transformation 
\beq
 \phi_1 \to e^{i\beta} \phi_1 ,\quad 
 \phi_2 \to e^{-i\beta} \phi_2 
\eeq
is explicitly broken by $\gamma \neq 0$.

Let us take strong coupling limit (with keeping $\gamma$ finite):
\beq
 e, \lambda_i \to \infty,
\eeq
giving  constraints
\beq
 |\phi_i| =1, \quad 
 \phi_i = e^{i\theta_i} .
\eeq
With taking a gauge
$A_{\mu} = \del_{\mu} \theta_2$ and 
defining the phase difference 
$\theta(x) \equiv \theta_1 (x) - \theta_2 (x)$, 
the covariant derivative terms in Lagrangian in 
Eq.~(\ref{eq:2compU(1)}) become
\beq
&& D_{\mu} \phi_1 
= i (\del_{\mu} \theta_1 - A_{\mu}) e^{i \theta_1} 
= i \del_{\mu} (\theta_1 - \theta_2) e^{i \theta_1} 
= i \del_{\mu} \theta e^{i \theta_1} , \non
&& D_{\mu} \phi_2 =  i (\del_{\mu} \theta_2 - A_{\mu}) e^{i \theta_2} = 0,
\eeq
while the Josephson terms in Eq.~(\ref{eq:Josephson}) become
\beq 
{\cal L}_{\rm J,1} 
= -m^2(1- \cos \theta) , \quad
{\cal L}_{\rm J,2} 
= -m^2(1- \cos^2 \theta) , 
\quad
 \gamma \equiv m^2.
\eeq
The gauge theory Lagrangian in Eq.~(\ref{eq:2compU(1)}) 
reduces the sine-Gordon model in  Eq.~(\ref{eq:SG}) 
or the modified sine-Gordon model in  Eq.~(\ref{eq:SG2}).

Let us remark on physical realizations of this model and 
its sine-Gordon solitons.
A non-relativistic version of the Lagrangian has 
the kinetic and gradient terms 
\beq
 \1{2} \sum_i (i \phi_i^* D_0 \phi_i  +{\rm h.c} 
 - D_a  \phi_i^*  D_a \phi_i)  . \label{eq:nonrela}
\eeq
instead of the first term in the Lagrangian in Eq.~(\ref{eq:2compU(1)}).
Here $a=1,2,(3)$ is a spatial index.
The linear Josephson term ${\cal L}_{\rm J,1}$ in Eq.~(\ref{eq:Josephson})
is relevant for the Landau-Ginzburg description of 
two-gap superconductors such as MgB$_2$, 
in which the term proportional to $\gamma$ is called the 
(internal) Josephson coupling and $\theta(x)$ is called the Leggett mode.
The sine-Gordon soliton is called the phase soliton in this context, 
which was first pointed out theoretically \cite{Tanaka:2001} 
and was found experimentally \cite{Gurevich:2003}.
It is also relevant for a Josephson junction 
of two superconductors.
On the other hand, the case with the quadratic Josephson interaction 
${\cal L}_{\rm J,2}$ in Eq.~(\ref{eq:Josephson})
is relevant for chiral p-wave superconductors \cite{Agterberg1998}, 
such as  Sr$_2$RuO$_4$.

A non-relativistic version of the Lagrangian 
(\ref{eq:nonrela})
in 
which overall $U(1)$ is not gauged ($e= 0$)
yields the Gross-Pitaevskii equation 
for two-component Bose-Einstein condensates 
of ultracold atomic gases such as Rb$_{87}$,  
in which the term proportional to $\gamma$ is called 
a Rabi oscillation term. 
(In addition, the term $g_{12}|\phi_1|^2|\phi_2|^2$ is also present 
but it is not important for the phase solitons.)
The sine-Gordon (phase) soliton in this case was studied in Ref.~\cite{Son:2001td}.

\subsection{Non-Abelian gauge theory}

Let us consider a $U(N)$ gauge theory 
coupled with two $N \times N$ matrix-valued 
complex scalar fields $\Phi_i(x)$ ($i=1,2$), whose Lagrangian is given by
\beq
 {\cal L} &=& \1{2} \sum_{i=1,2} \tr D_{\mu}\Phi_i^\dagger D^{\mu}\Phi_i
 + {\gamma\over 2} \tr (\Phi_1^\dagger \Phi_2 + {\rm h.c.}-2{\bf 1}_N) \non
 &&- \sum_{i=1,2} {\lambda_i \over 4}  \tr (\Phi_i^\dagger\Phi_i -{\bf 1}_N)^2
 + \1{4 g^2} \tr F_{\mu\nu}^2 \label{eq:2compU(N)}
\eeq
with $D_{\mu} \Phi_i = (\del_{\mu} - i A_{\mu} )\Phi_i$ 
and $A_{\mu} = A_{\mu}^A(x) T_A$ with 
$U(N)$ generators $T_A$.
${\cal L}_{\rm J}$ is a non-Abelian Josephson term either linear 
or quadratic: 
\beq
&&  {\cal L}_{\rm J,1} = {\gamma\over 2} \tr (\Phi_1^\dagger \Phi_2 + {\rm h.c.}-2{\bf 1}_N) ,\non
&&  {\cal L}_{\rm J,2} = {\gamma\over 2} \tr \left[(\Phi_1^\dagger \Phi_2)^2 + {\rm h.c.}-2{\bf 1}_N \right] .
\label{eq:NA-Josephson}
\eeq
The $U(N)_{\rm V}$ gauge transformation is defined by
\beq
&& \Phi_i \to V(x)\Phi_i, \quad 
 A_{\mu} \to V(x)A_{\mu}  V(x)^{-1}+ i V(x)\del_{\mu} V^{-1}(x), 
\eeq
while a $U(N)_{\rm A}$ global transformation 
\beq
 \Phi_1 \to g \Phi_1 ,\quad 
 \Phi_2 \to g^{-1} \Phi_2 ,\quad g \in U(N)_{\rm A}
\eeq
is explicitly broken by $\gamma \neq 0$.

Let us take strong coupling limit  (with keeping $\gamma$ finite):
\beq
 g, \lambda_i \to \infty,
\eeq
giving  constraints
\beq
 \Phi_i^\dagger \Phi_i ={\bf 1}_N . 
\eeq
These constraints can be solved as 
\beq
 \Phi_1 (x) = \hat U(x) , \quad 
 \Phi_2 (x)= \hat U^\dagger (x),\quad 
 \hat U(x) \in U(N).
\eeq
With taking a gauge 
$A_{\mu} = i \hat U^\dagger \del_{\mu} \hat U$  
and defining 
$U(x) \equiv \hat U^2(x)$, 
the covariant derivative terms in Lagrangian in 
Eq.~(\ref{eq:2compU(N)}) become
\beq
&& D_{\mu} \Phi_1 
= \del_{\mu}\hat U - i A_{\mu}\hat U 
= \del_{\mu} U(x), \quad
 D_{\mu} \Phi_2 
= \del_{\mu} \hat U^\dagger - i A_{\mu} \hat U^\dagger
= 0, 
\eeq
and the Josephson terms reduce to
\beq
&& {\cal L}_{\rm J,1} 
=  -m^2 \tr (2{\bf 1}_N- U - U^\dagger) , \non
&& {\cal L}_{\rm J,2} 
=  -m^2 \tr (2{\bf 1}_N- U^2 - U^{\dagger 2} ) , \\
&& \gamma \equiv m^2. \nonumber
\eeq
Therefore, the gauge theory Lagrangian in Eq.~(\ref{eq:2compU(N)}) 
reduces the non-Abelian sine-Gordon model in  Eq.~(\ref{eq:U(N)SG}) 
or the modified non-Abelian  sine-Gordon model in  Eq.~(\ref{eq:SG2}).

The relativistic Lagrangian in Eq.~(\ref{eq:2compU(N)}) 
is relevant for 
a linear model description of chiral Lagrangian using 
a hidden local gauge symmetry 
for which gauge bosons of $U(N)$ gauge symmetry is 
vector mesons of the hidden local symmetry, 
see, e.~g.~Ref.~\cite{Kitano:2012zz}.

A non-relativistic version of the Lagrangian has 
the kinetic and gradient terms 
\beq
  \1{2} \sum_i \tr (i \Phi_i^\dagger D_0 \Phi_i  +{\rm h.c} 
 - D_a  \Phi_i^*  D_a \Phi_i)   \label{eq:nonrela-NA}
\eeq
instead of the first term in the Lagrangian in Eq.~(\ref{eq:2compU(N)}). 
The non-relativistic case with $N=3$ with ungauged $U(1)$  
is relevant for the Landau-Ginzburg description of 
the color-flavor locking phase (a color superconductor) for high density QCD \cite{Alford:2001dt,Eto:2013hoa}.
In this case, 
$(\Phi_1)_{\alpha i} 
=  \epsilon_{\alpha\beta\gamma}  \epsilon_{ijk}
q^L_{j\beta} q^L_{k\gamma}$ 
and $(\Phi_2)_{\alpha i} 
=  \epsilon_{\alpha\beta\gamma}  \epsilon_{ijk}
q^R_{j\beta} q^R_{k\gamma}$ 
are diquark condensates of left and right handed quarks 
$q^L_{j\beta}$ and $q^R_{j\beta}$, respectively,  
where $\alpha,\beta,\gamma=1,2,3$ and
$i,j,k=1,2,3$ are color and flavor indices, respectively.

Here, we have considered the potential for the $U(1)$ symmetry 
induced from quark mass in chiral Lagrangian in QCD.
On the other hand, 
there is another potential term $V \sim \det \Phi_1 + \det \Phi_2$
induced from the $U(1)_{\rm A}$ anomaly at quantum level.
The non-Abelian sine-Gordon kink should be deformed by this potential 
accordingly \cite{Eto:2013bxa}. 
Therefore, in real QCD, 
our solutions are relevant in asymptotically 
high temperature or high density, in which  
the $U(1)_{\rm A}$ anomaly disappears.

\section{Non-Abelian Vortex that Terminates Non-Abelian 
Sine-Gordon Kink} \label{sec:NA-vortex}

The $U(N)$ chiral Lagrangian or more precisely 
the corresponding $U(N)$ linear sigma model admits 
a non-Abelian global vortex 
\cite{Balachandran:2002je,Nitta:2007dp,Nakano:2007dq,Eto:2009wu},
see Ref.~\cite{Eto:2013hoa} as a review. 
When one discusses the asymptotic form of the vortex solution, 
the chiral Lagrangian is enough.
Here, we briefly discuss a relation between 
the non-Abelian global vortex and 
the non-Abelian sine-Gordon kink.

Let $(r,\ph,z)$ be cylindrical coordinates of space.
Then, the asymptotic form of a non-Abelian global vortex 
can be written as
\beq
 U(r \to \infty ,\ph,z) = {\rm diag} (e^{i \theta(\ph)},1,\cdots,1). \label{eq:ansatz-vor}
\eeq
In the limit of no mass term ($m=0$), the unit winding solution 
is simply given by $\theta=\ph$ so that 
the vortex is axisymmetric.
The configuration in Eq.~(\ref{eq:ansatz-vor}) can be rewritten as
\beq
  U(r \to \infty ,\ph,z) 
 &=& \exp \left(i {\theta(\ph) \over N}\right) 
 \exp \left(i \theta(x) T_0 \right),\label{eq:T_0-vor}\\
T_0 &\equiv& {1 \over N} {\rm diag.}(N-1,-1,\cdots,-1) .
\nonumber
\eeq
It is obvious that the configuration of the vortex 
breaks the $SU(N)_{\rm V}$ symmetry of the vacuum 
to a subgroup $SU(N-1)\times U(1)$ 
so that there appear moduli ${\mathbb C}P^{N-1}$,  
although these moduli are non-normalizable
\cite{Nitta:2007dp,Nakano:2007dq}.

\begin{figure}
\begin{center}
\includegraphics[width=0.40\linewidth,keepaspectratio]{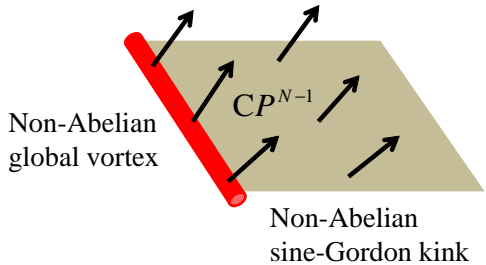} 
\end{center}
\caption{A junction of a non-Abelian vortex and 
a non-Abelian sine-Gordon domain wall.
A non-Abelian vortex is attached by a 
non-Abelian sine-Gordon kink in the presence of the mass. 
In other words, the latter can terminate on the former.
The ${\mathbb C}P^{N-1}$ moduli, that are denoted by arrows, 
match at the junction line. 
\label{fig:SG-kink-vortex}}
\end{figure}

In the presence of the mass term ($m \neq 0$), 
the global vortex configuration is deformed and is 
no more axisymmetric. 
In this case, the potential term appears 
for the field $\theta(\ph)$ in the vortex ansatz in 
Eq.~(\ref{eq:ansatz-vor}). 
This is of course the sine-Gordon potential discussed 
in the previous sections. 
Only the difference is the argument of $\theta$ is 
$\theta(\ph)$ here and $\theta(x)$ before.
The final configuration is 
a non-Abelian vortex attached by a 
non-Abelian sine-Gordon kink,
as schematically drawn in Fig.~\ref{fig:SG-kink-vortex}. 
Both the non-Abelian vortex and non-Abelian sine-Gordon kink 
have the ${\mathbb C}P^{N-1}$ moduli, 
and consequently they match at a junction line \footnote{
This situation is the same with the case of 
a non-Abelian monopole that terminates a non-Abelian 
vortex string \cite{Auzzi:2003em,Eto:2006dx}. 
This not only because 
 the both ${\mathbb C}P^{N-1}$ moduli  of 
the monopole and vortex match at junction point 
but also because the monopole moduli are 
non-normalizable.
}.

\begin{figure}
\begin{center}
\begin{tabular}{cc}
\includegraphics[width=0.40\linewidth,keepaspectratio]{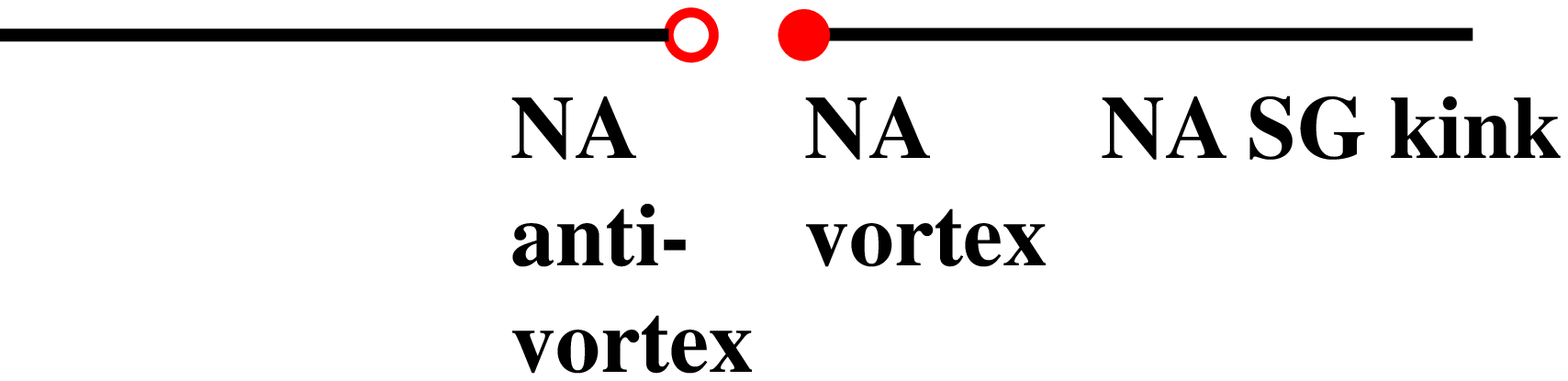} 
&
\includegraphics[width=0.40\linewidth,keepaspectratio]{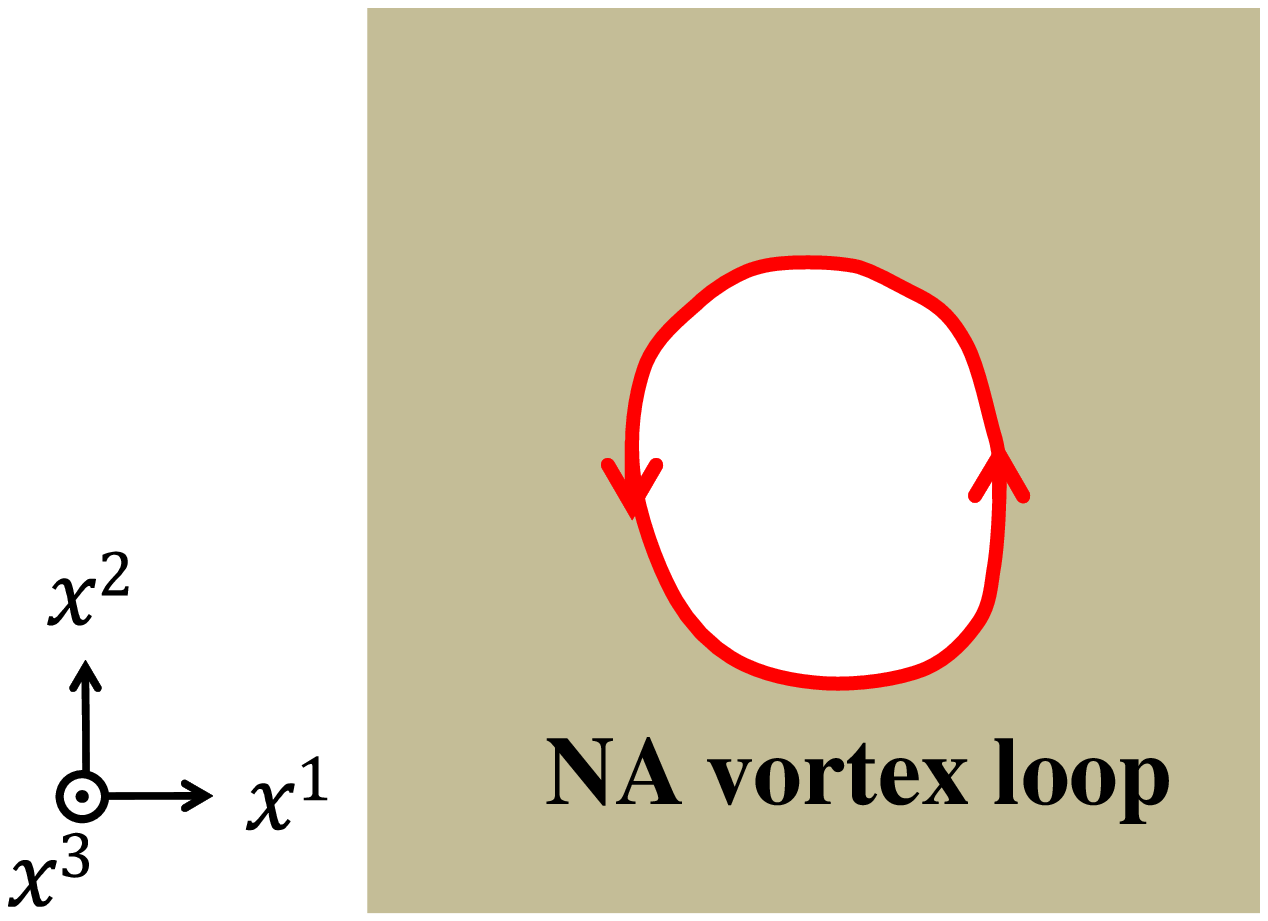} \\
(a) & (b)
\end{tabular}
\end{center}
\caption{Decay of a non-Abelian sine-Gordon kink. 
(a) In $d=2+1$, a non-Abelian sine-Gordon domain line can decay 
by creating a pair of a non-Abelian vortex and a non-Abelian anti-vortex. 
(b) In $d=3+1$ the non-Abelian sine-Gordon domain wall 
can decay by creating a hole bound by a closed non-Abelian vortex string.
These processes can occur either thermally or by quantum tunneling. 
\label{fig:SG-kink-decay}}
\end{figure}

This fact implies the instability of sine-Gordon kinks 
in the $U(N)$ {\it linear} sigma model 
as in the same manner with an axion string 
\cite{Vilenkin:2000}.
In $d=2+1$, the sine-Gordon domain line can terminate on 
a global non-Abelian vortex 
\cite{Eto:2013hoa,Balachandran:2002je,Nitta:2007dp,Nakano:2007dq,Eto:2009wu}.
The domain line can decay 
by creating a pair of a non-Abelian vortex and a non-Abelian anti-vortex,
as shown in Fig.~\ref{fig:SG-kink-decay} (a).
In $d=3+1$, the non-Abelian sine-Gordon domain wall 
can decay by creating a hole bound by a closed non-Abelian vortex string, 
as illustrated in Fig.~\ref{fig:SG-kink-decay} (b).  
This process can occur either thermally or by quantum tunneling. 
More details will be discussed elsewhere.
However, note that the instability does not exist 
in the nonlinear model, the $U(N)$ chiral Lagrangian. 
This is the same situation with an axion string \cite{Vilenkin:2000}.

\section{Summary and Discussion}\label{sec:summary}

We have pointed out that the $U(N)$ chiral Lagrangian admits 
a non-Abelian sine-Gordon kink 
that carries non-Abelian moduli  
${\mathbb C}P^{N-1} \simeq SU(N)/[SU(N-1)\times U(1)]$.
We have also presented the non-Abelian gauge theory 
that admits the same non-Abelian sine-Gordon kink.
In the Abelian case, this reduces to the Lagrangian for 
two-gap superconductors.
Two possibilities to realize it in QCD have been discussed.
We have also briefly discussed in the $U(N)$ linear sigma model 
that a sine-Gordon kink can terminate on a 
non-Abelian global vortex, implying the instability of 
the sine-Gordon kink in the linear model.

Several discussions are addressed here.
One of the most important task remaining 
is constructing the low-energy effective theory 
by the moduli approximation \cite{Manton:1981mp},
which is the ${\mathbb C}P^{N-1}$ model.
One then can construct 
${\mathbb C}P^{N-1}$ lumps on it 
that would represent $U(N)$ Skyrmions 
as was so for the $SU(2)$ model with two vacua 
\cite{Nitta:2012wi,Gudnason:2014nba}.
See Ref.~\cite{Eto:2015uqa} for a further study along this line.

The interaction between two kinks
located at $x=X_{1,2}$ with the orientations $\phi_{1,2}$ 
can be considered. 
Like the Abrikosov-type ansatz for vortices, 
we can give an ansatz for the total configuration as
$U_{\rm tot}(x) = U_1(x-X_1,\phi_1) U_2(x-X_2,\phi_2)$ 
for well-separated kinks $|X_1-X_2| >> m^{-1}$.
In particular, 
an Abelian sine-Gordon kink 
would be separated into $N$ non-Abelian kinks 
without cost of energy, 
which can be expected from the fact that 
an Abelian kink has energy $N$ multiple of 
those of non-Abelian kinks.  
A similar calculation was done for 
the force between two non-Abelian global vortices 
\cite{Nakano:2007dq,Eto:2013hoa}.

In two-gap superconductors, 
a unit winding vortex can be split into 
two fractional vortices winding around different components, 
which are connected by a sine-Gordon kink 
\cite{Babaev:2001hv,Goryo:2007,Nitta:2010yf}.
The same happens for 
coherently coupled multi-component BECs
\cite{Kasamatsu:2004,Cipriani:2013nya,Eto:2012rc}.
In the same way,
a local non-Abelian vortex can be split into 
a set of two global non-Abelian vortices connected 
by a non-Abelian sine-Gordon domain wall discussed here. 
In the case of the color-flavor locked phase of dense quark matter, 
a non-Abelian vortex \cite{Balachandran:2005ev,Eto:2013hoa} 
has 1/3 fractional $U(1)$ winding 
in both $\Phi_1$ and $\Phi_2$,
but it may be decomposed into 
a global vortex with 1/6 $U(1)$ winding (1/3 $U(1)$ 
winding in only one of  $\Phi_1$ and $\Phi_2$). 
This will be also discussed elsewhere.

The $U(N)$ principal chiral model studied in this paper has been found to appear as the effective theory of a non-Abelian domain wall \cite{Shifman:2003uh}. If a Josephson term is added in the bulk theory, this domain wall behaves as a Josephson junction of two color superconductors and the mass term is induced in the $U(N)$ principal chiral model on the wall \cite{Nitta:2015mma}. Then, non-Abelian sine-Gordon solitons describe non-Abelian Josephson vortices, that is, non-Abelian vortices trapped inside the Josephson junction \cite{Nitta:2015mma,Nitta:2015mxa}.

Non-Abelian $U(N)$ Sine-Gordon kinks can be extended to the case of arbitrary gauge groups $G$ in the form of 
${G \times U(1) \over {\mathbb Z}_r}$ 
with the center ${\mathbb Z}_r$ of $G$, 
since non-Abelian vortices with this type of gauge groups 
were studied before \cite{Eto:2008yi}, 
such as $SO(N)$ and $USp(2N)$ groups \cite{Eto:2009bg}.

Finally, the sine-Gordon model is integrable. Therefore, 
we expect the non-Abelian sine-Gordon model 
presented here is also integrable.

\section*{Acknowledgments}

We thank Minoru Eto for useful comments. 
This work is supported in part by Grant-in-Aid for Scientific Research 
No.~25400268
and by the ``Topological Quantum Phenomena'' 
Grant-in-Aid for Scientific Research 
on Innovative Areas (No.~25103720)  
from the Ministry of Education, Culture, Sports, Science and Technology 
(MEXT) of Japan.


\newcommand{\J}[4]{{\sl #1} {\bf #2} (#3) #4}
\newcommand{\andJ}[3]{{\bf #1} (#2) #3}
\newcommand{\AP}{Ann.\ Phys.\ (N.Y.)}
\newcommand{\MPL}{Mod.\ Phys.\ Lett.}
\newcommand{\NP}{Nucl.\ Phys.}
\newcommand{\PL}{Phys.\ Lett.}
\newcommand{\PR}{ Phys.\ Rev.}
\newcommand{\PRL}{Phys.\ Rev.\ Lett.}
\newcommand{\PTP}{Prog.\ Theor.\ Phys.}
\newcommand{\hep}[1]{{\tt hep-th/{#1}}}


\begin{thebibliography}{100}
\bibitem{Perring:1962vs} 
  J.~K.~Perring and T.~H.~R.~Skyrme,
  ``A Model unified field equation,''
  Nucl.\ Phys.\  {\bf 31}, 550 (1962).

\bibitem{Manton:2004tk} 
  N.~S.~Manton and P.~Sutcliffe,
  ``Topological solitons,''
  Cambridge, UK: Univ. Pr. (2004) 493 p.

\bibitem{Rajaraman:1982is} 
  R.~Rajaraman,
  ``Solitons And Instantons. An Introduction To Solitons And Instantons In Quantum Field Theory,''
  Amsterdam, Netherlands: North-holland ( 1982) 409p.


\bibitem{Eto:2013hoa} 
  M.~Eto, Y.~Hirono, M.~Nitta and S.~Yasui,
  ``Vortices and Other Topological Solitons in Dense Quark Matter,''
  PTEP {\bf 2014}, no. 1, 012D01 (2014)
  [arXiv:1308.1535 [hep-ph]].

\bibitem{Vilenkin:2000}
A.~Vilenkin and E.~P.~S.~Shellard, 
{\it Cosmic Strings and Other Topological Defects}, (Cambridge Monographs on Mathematical Physics), Cambridge University Press (July 31, 2000).


\bibitem{Ustinov:1998}
A.~V.~Ustinov,
``Solitons in Josephson junctions,"
Physica D {\bf 123}, 315–329 (1998). 

\bibitem{Blatter:1994}
G.~Blatter, M.~V.~Feigel'man, V.~ B.~Geshkenbein, A.~I.~Larkin, 
and V.~M.~Vinokur,
``Vortices in high-temperature superconductors," 
Rev.\ Mod.\ Phys.\ {\bf 66}, 1125–1388 (1994) 

\bibitem{Tanaka:2001}
Y.~Tanaka,
``Phase Instability in Multi-band Superconductors," 
J.\ Phys.\ Soc.\ Jp.\ {\bf 70} (2001)  2844;  
``Soliton in Two-Band Superconductor,"
Phys.\ Rev.\ Lett.\ {\bf 88},  017002 (2001).


\bibitem{Gurevich:2003}
A.~Gurevich and V.~M.~Vinokur,
``Interband Phase Modes and Nonequilibrium Soliton Structures in Two-Gap Superconductors," 
Phys.\ Rev.\ Lett.\ {\bf 90}, 047004 (2003).


\bibitem{Goryo:2007}
J. Goryo, S. Soma and H. Matsukawa, 
``Deconfinement of vortices with continuously variable fractions of the unit flux quanta in two-gap superconductors,"
 Euro Phys.\ Lett.\ {\bf 80}, 17002 (2007) 
 [arXiv:cond-mat/0608015].

\bibitem{Agterberg1998}
D.~F.~Agterberg, 
``Vortex Lattice Structures of Sr$_2$RuO$_4$,"
Phys.\ Rev.\ Lett.\ {\bf 80}, 5184 (1998); 
J.~Garaud and E.~Babaev, 
``Skyrmionic state and stable half-quantum vortices in chiral p-wave superconductors,"
Phys. Rev. B 86, 060514 (2012). 

\bibitem{Son:2001td} 
  D.~T.~Son and M.~A.~Stephanov,
  ``Domain walls in two-component Bose-Einstein condensates,''  
Phys.\ Rev.\ A {\bf 65}, 063621 (2002)  [cond-mat/0103451].  

\bibitem{Kaurov:2005}
V.~M.~Kaurov and A.~B.~Kuklov, 
Phys.\ Rev.\ A {\bf 71}, 011601 (2005)
[arXiv:cond-mat/0312084]; 
``Atomic Josephson vortices,"
Phys.\ Rev.\ A {\bf 73}, 013627 (2006) 
[arXiv:cond-mat/0508342 [cond-mat.other]].

\bibitem{Volovik2003}
G.~E.~Volovik,
{\it The Universe in a Helium Droplet}, 
Clarendon Press,  Oxford (2003).



\bibitem{Chen:1977}
C.~W.~Chen,  
``Magnetism and metallurgy of soft magnetic materials,"
Dover Pubns (1977); 
A.~P.~Malozemoff,  J.~C.~Slonczewski, 
``Magnetic domain walls in bubble materials," 
Academic Press  (New York) (1979).

\bibitem{Nitta:2012xq} 
  M.~Nitta,
  ``Josephson vortices and the Atiyah-Manton construction,''
  Phys.\ Rev.\ D {\bf 86}, 125004 (2012)
  [arXiv:1207.6958 [hep-th]].

\bibitem{Babaev:2001hv} 
  E.~Babaev,
  ``Vortices carrying an arbitrary fraction of magnetic flux quantum in two gap superconductors,''
  Phys.\ Rev.\ Lett.\  {\bf 89}, 067001 (2002)
  [cond-mat/0111192];
  E.~Babaev,
  ``Phase diagram of a planar two band superconductor: Condensation of vortices with fractional flux quantum and existence of a nonsuperconducting superfluid state in this system,''
  Nucl.\ Phys.\ B {\bf 686}, 397 (2004)
  [cond-mat/0201547];
  J.~Smiseth, E.~Smorgrav, E.~Babaev and A.~Sudbo,
  ``Field and temperature induced topological phase transitions in the three-dimensional N-component London superconductor,''
  Phys.\ Rev.\ B {\bf 71}, 214509 (2005)
  [cond-mat/0411761].


\bibitem{Nitta:2010yf} 
  M.~Nitta, M.~Eto, T.~Fujimori and K.~Ohashi,
  ``Baryonic Bound State of Vortices in Multicomponent Superconductors,''  J.\ Phys.\ Soc.\ Jap.\  {\bf 81}, 084711 (2012)  [arXiv:1011.2552 [cond-mat.supr-con]].  

\bibitem{Kasamatsu:2004}
K.~Kasamatsu, M.~Tsubota and M.~Ueda, 
``Vortex Molecules in 
Coherently Coupled Two-Component Bose-Einstein Condensates,"
Phys. Rev. Lett. {\bf 93}, 250406 (2004);

\bibitem{Cipriani:2013nya} 
  M.~Cipriani and M.~Nitta,
  ``Crossover between integer and fractional vortex lattices in coherently coupled two-component Bose-Einstein condensates,''
  Phys.\ Rev.\ Lett.\  {\bf 111}, 170401 (2013)
  [arXiv:1303.2592 [cond-mat.quant-gas]].


\bibitem{Eto:2012rc} 
  M.~Eto and M.~Nitta,
  ``Vortex trimer in three-component Bose-Einstein condensates,''  Phys.\ Rev.\ A {\bf 85}, 053645 (2012)  [arXiv:1201.0343 [cond-mat.quant-gas]].  
  M.~Eto and M.~Nitta,
  ``Vortex graphs as N-omers and CP(N-1) Skyrmions in N-component Bose-Einstein condensates,''
  Europhys.\ Lett.\  {\bf 103}, 60006 (2013)
  [arXiv:1303.6048 [cond-mat.quant-gas]];
  M.~Nitta, M.~Eto and M.~Cipriani,
  ``Vortex molecules in Bose-Einstein condensates,''
  J.\ Low.\ Temp.\ Phys.\  {\bf 175}, 177 (2013)
  [arXiv:1307.4312 [cond-mat.quant-gas]].



\bibitem{Auzzi:2006ju} 
  R.~Auzzi, M.~Shifman and A.~Yung,
  ``Domain Lines as Fractional Strings,''  
Phys.\ Rev.\ D {\bf 74}, 045007 (2006)  [hep-th/0606060].  

\bibitem{Kobayashi:2013ju} 
  M.~Kobayashi and M.~Nitta,
  ``Jewels on a wall ring,''
  Phys.\ Rev.\ D {\bf 87}, 085003 (2013)
  [arXiv:1302.0989 [hep-th]].

\bibitem{Jennings:2013aea} 
  P.~Jennings and P.~Sutcliffe,
  ``The dynamics of domain wall Skyrmions,''
  J.\ Phys.\ A {\bf 46}, 465401 (2013)
  [arXiv:1305.2869 [hep-th]].

\bibitem{Sutcliffe:1992ep}
  P.~M.~Sutcliffe,
  ``Sine-Gordon Solitons From Cp(1) Instantons,''
  Phys.\ Lett.\ B {\bf 283}, 85 (1992).

\bibitem{Stratopoulos:1992hq}
  G.~N.~Stratopoulos and W.~J.~Zakrzewski,
  ``Approximate Sine-Gordon solitons,''
  Z.\ Phys.\ C {\bf 59}, 307 (1993).

\bibitem{Nitta:2012wi} 
  M.~Nitta,
  ``Correspondence between Skyrmions in 2+1 and 3+1 Dimensions,''
  Phys.\ Rev.\ D {\bf 87}, 025013 (2013)
  [arXiv:1210.2233 [hep-th]].

\bibitem{Gudnason:2014nba} 
  S.~B.~Gudnason and M.~Nitta,
  ``Domain wall Skyrmions,''
  Phys.\ Rev.\ D {\bf 89}, 085022 (2014)
  [arXiv:1403.1245 [hep-th]].

\bibitem{Nitta:2012rq} 
  M.~Nitta,
  ``Matryoshka Skyrmions,''
  Nucl.\ Phys.\ B {\bf 872}, 62 (2013)
  [arXiv:1211.4916 [hep-th]].



\bibitem{Gudnason:2014hsa} 
  S.~B.~Gudnason and M.~Nitta,
  ``Incarnations of Skyrmions,''
  Phys.\ Rev.\ D {\bf 90}, 085007 (2014)
  [arXiv:1407.7210 [hep-th]];
  S.~B.~Gudnason and M.~Nitta,
  ``Baryonic torii: Toroidal baryons in a generalized Skyrme model,''
  Phys.\ Rev.\ D {\bf 91}, no. 4, 045027 (2015)
  [arXiv:1410.8407 [hep-th]].

\bibitem{Gudnason:2014gla} 
  S.~B.~Gudnason and M.~Nitta,
  ``Effective field theories on solitons of generic shapes,''
  arXiv:1407.2822 [hep-th].

\bibitem{Kobayashi:2013xoa} 
  M.~Kobayashi and M.~Nitta,
  ``Torus knots as Hopfions,''
  Phys.\ Lett.\ B {\bf 728}, 314 (2014)
  [arXiv:1304.6021 [hep-th]];
  M.~Kobayashi and M.~Nitta,
  ``Toroidal domain walls as Hopfions,''
  arXiv:1304.4737 [hep-th].

\bibitem{Nitta:2013cn} 
  M.~Nitta,
  ``Instantons confined by monopole strings,''
  Phys.\ Rev.\ D {\bf 87}, no. 6, 066008 (2013)
  [arXiv:1301.3268 [hep-th]];
  M.~Nitta,
  ``Incarnations of Instantons,''
  Nucl.\ Phys.\ B {\bf 885}, 493 (2014)
  [arXiv:1311.2718 [hep-th]].


\bibitem{Lund:1976ze} 
  F.~Lund and T.~Regge,
  ``Unified Approach to Strings and Vortices with Soliton Solutions,''
  Phys.\ Rev.\ D {\bf 14}, 1524 (1976).

\bibitem{Pohlmeyer:1975nb} 
  K.~Pohlmeyer,
  ``Integrable Hamiltonian Systems and Interactions Through Quadratic Constraints,''
  Commun.\ Math.\ Phys.\  {\bf 46}, 207 (1976).

\bibitem{Bakas:1993xh} 
  I.~Bakas,
  ``Conservation laws and geometry of perturbed coset models,''
  Int.\ J.\ Mod.\ Phys.\ A {\bf 9}, 3443 (1994)
  [hep-th/9310122].

\bibitem{Gepner:1984au} 
  D.~Gepner,
  ``Nonabelian Bosonization and Multiflavor {QED} and {QCD} in Two-dimensions,''
  Nucl.\ Phys.\ B {\bf 252}, 481 (1985).


\bibitem{Naganuma:2001br} 
  M.~Naganuma, M.~Nitta and N.~Sakai,
  ``BPS walls and junctions in SUSY nonlinear sigma models,''
  Phys.\ Rev.\ D {\bf 65}, 045016 (2002)
  [hep-th/0108179].


\bibitem{Park:1995rj} 
  Q.~H.~Park and H.~J.~Shin,
  ``Classical matrix sine-Gordon theory,''
  Nucl.\ Phys.\ B {\bf 458}, 327 (1996)
  [hep-th/9505017].

\bibitem{Bakas:1995bm} 
  I.~Bakas, Q.~H.~Park and H.~J.~Shin,
  ``Lagrangian formulation of symmetric space sine-Gordon models,''
  Phys.\ Lett.\ B {\bf 372}, 45 (1996)
  [hep-th/9512030].




\bibitem{Hanany:2003hp} 
  A.~Hanany and D.~Tong,
  ``Vortices, instantons and branes,''
  JHEP {\bf 0307}, 037 (2003)
  [hep-th/0306150].

\bibitem{Auzzi:2003fs} 
  R.~Auzzi, S.~Bolognesi, J.~Evslin, K.~Konishi and A.~Yung,
  ``NonAbelian superconductors: Vortices and confinement in N=2 SQCD,''
  Nucl.\ Phys.\ B {\bf 673}, 187 (2003)
  [hep-th/0307287].

\bibitem{Shifman:2004dr} 
  M.~Shifman and A.~Yung,
  ``NonAbelian string junctions as confined monopoles,''  Phys.\ Rev.\ D {\bf 70}, 045004 (2004)  [hep-th/0403149];  
  A.~Hanany and D.~Tong,
  ``Vortex strings and four-dimensional gauge dynamics,''  JHEP {\bf 0404}, 066 (2004)  [hep-th/0403158].  
  M.~Eto, Y.~Isozumi, M.~Nitta, K.~Ohashi and N.~Sakai,
  ``Instantons in the Higgs phase,''
  Phys.\ Rev.\ D {\bf 72}, 025011 (2005)
  [hep-th/0412048].


\bibitem{Eto:2005yh} 
  M.~Eto, Y.~Isozumi, M.~Nitta, K.~Ohashi and N.~Sakai,
  ``Moduli space of non-Abelian vortices,''
  Phys.\ Rev.\ Lett.\  {\bf 96}, 161601 (2006)
  [hep-th/0511088];
  M.~Eto, K.~Konishi, G.~Marmorini, M.~Nitta, K.~Ohashi, W.~Vinci and N.~Yokoi,
  ``Non-Abelian Vortices of Higher Winding Numbers,''
  Phys.\ Rev.\ D {\bf 74}, 065021 (2006)
  [hep-th/0607070];
  M.~Eto, K.~Hashimoto, G.~Marmorini, M.~Nitta, K.~Ohashi and W.~Vinci,
  ``Universal Reconnection of Non-Abelian Cosmic Strings,''
  Phys.\ Rev.\ Lett.\  {\bf 98}, 091602 (2007)
  [hep-th/0609214].

\bibitem{Tong:2005un} 
  D.~Tong,
  ``TASI lectures on solitons: Instantons, monopoles, vortices and kinks,''
  hep-th/0509216.

\bibitem{Eto:2006pg} 
  M.~Eto, Y.~Isozumi, M.~Nitta, K.~Ohashi and N.~Sakai,
  ``Solitons in the Higgs phase: The Moduli matrix approach,''
  J.\ Phys.\ A {\bf 39}, R315 (2006)
  [hep-th/0602170].

\bibitem{Shifman:2007ce} 
  M.~Shifman and A.~Yung,
  ``Supersymmetric Solitons and How They Help Us Understand Non-Abelian Gauge Theories,''
  Rev.\ Mod.\ Phys.\  {\bf 79}, 1139 (2007)
  [hep-th/0703267];
  M.~Shifman and A.~Yung,
  ``Supersymmetric solitons,''
  Cambridge, UK: Cambridge Univ. Pr. (2009) 259 p

\bibitem{Auzzi:2003em} 
  R.~Auzzi, S.~Bolognesi, J.~Evslin and K.~Konishi,
  ``Non-Abelian monopoles and the vortices that confine them,''
  Nucl.\ Phys.\ B {\bf 686}, 119 (2004)
  [hep-th/0312233].

\bibitem{Eto:2006dx} 
  M.~Eto, L.~Ferretti, K.~Konishi, G.~Marmorini, M.~Nitta, K.~Ohashi, W.~Vinci and N.~Yokoi,
  ``Non-Abelian duality from vortex moduli: A Dual model of color-confinement,''
  Nucl.\ Phys.\ B {\bf 780}, 161 (2007)
  [hep-th/0611313].


\bibitem{Balachandran:2002je} 
  A.~P.~Balachandran and S.~Digal,
  ``NonAbelian topological strings and metastable states in linear sigma model,''
  Phys.\ Rev.\ D {\bf 66}, 034018 (2002)
  [hep-ph/0204262].

\bibitem{Nitta:2007dp} 
  M.~Nitta and N.~Shiiki,
  ``Non-Abelian Global Strings at Chiral Phase Transition,''
  Phys.\ Lett.\ B {\bf 658}, 143 (2008)
  [arXiv:0708.4091 [hep-ph]].

\bibitem{Nakano:2007dq} 
  E.~Nakano, M.~Nitta and T.~Matsuura,
  ``Interactions of non-Abelian global strings,''  Phys.\ Lett.\ B {\bf 672}, 61 (2009)  [arXiv:0708.4092 [hep-ph]];  
  E.~Nakano, M.~Nitta and T.~Matsuura,
  ``Non-Abelian Strings in Hot or Dense QCD,''
  Prog.\ Theor.\ Phys.\ Suppl.\  {\bf 174}, 254 (2008)
  [arXiv:0805.4539 [hep-ph]].
\bibitem{Eto:2009wu} 
  M.~Eto, E.~Nakano and M.~Nitta,
  ``Non-Abelian Global Vortices,''
  Nucl.\ Phys.\ B {\bf 821}, 129 (2009)
  [arXiv:0903.1528 [hep-ph]].


\bibitem{Alford:2001dt} 
  M.~G.~Alford,
  ``Color superconducting quark matter,''
  Ann.\ Rev.\ Nucl.\ Part.\ Sci.\  {\bf 51}, 131 (2001)
  [hep-ph/0102047].

\bibitem{Kitano:2012zz} 
  R.~Kitano, M.~Nakamura and N.~Yokoi,
  ``Making confining strings out of mesons,''
  Phys.\ Rev.\ D {\bf 86}, 014510 (2012)
  [arXiv:1202.3260 [hep-ph]].

\bibitem{Eto:2013bxa} 
  M.~Eto, Y.~Hirono and M.~Nitta,
  ``Domain Walls and Vortices in Chiral Symmetry Breaking,''
  PTEP {\bf 2014}, no. 3, 033B01 (2014)
  [arXiv:1309.4559 [hep-ph]].



\bibitem{Manton:1981mp} 
  N.~S.~Manton,
  ``A Remark on the Scattering of BPS Monopoles,''  Phys.\ Lett.\ B {\bf 110}, 54 (1982);  
  M.~Eto, Y.~Isozumi, M.~Nitta, K.~Ohashi and N.~Sakai,
  ``Manifestly supersymmetric effective Lagrangians on BPS solitons,''
  Phys.\ Rev.\ D {\bf 73}, 125008 (2006)
  [hep-th/0602289].

\bibitem{Eto:2015uqa} 
  M.~Eto and M.~Nitta,
  ``Non-Abelian Sine-Gordon Solitons: Correspondence between $SU(N)$ Skyrmions and ${\mathbb C}P^{N-1}$ Lumps,''
  Phys.\ Rev.\ D (in press) [arXiv:1501.07038 [hep-th]].

\bibitem{Balachandran:2005ev} 
  A.~P.~Balachandran, S.~Digal and T.~Matsuura,
  ``Semi-superfluid strings in high density QCD,''
  Phys.\ Rev.\ D {\bf 73}, 074009 (2006)
  [hep-ph/0509276];
  E.~Nakano, M.~Nitta and T.~Matsuura,
  ``Non-Abelian strings in high density QCD: Zero modes and interactions,''
  Phys.\ Rev.\ D {\bf 78}, 045002 (2008)
  [arXiv:0708.4096 [hep-ph]];
  M.~Eto and M.~Nitta,
  ``Color Magnetic Flux Tubes in Dense QCD,''
  Phys.\ Rev.\ D {\bf 80}, 125007 (2009)
  [arXiv:0907.1278 [hep-ph]];
  M.~Eto, E.~Nakano and M.~Nitta,
  ``Effective world-sheet theory of color magnetic flux tubes in dense QCD,''
  Phys.\ Rev.\ D {\bf 80}, 125011 (2009)
  [arXiv:0908.4470 [hep-ph]];
  M.~Eto, M.~Nitta and N.~Yamamoto,
  ``Instabilities of Non-Abelian Vortices in Dense QCD,''
  Phys.\ Rev.\ Lett.\  {\bf 104}, 161601 (2010)
  [arXiv:0912.1352 [hep-ph]].


\bibitem{Shifman:2003uh}
  M.~Shifman and A.~Yung,
  ``Localization of nonAbelian gauge fields on domain walls at weak
coupling (D-brane prototypes II),''
  Phys.\ Rev.\ D {\bf 70}, 025013 (2004)
  [hep-th/0312257];
  M.~Eto, M.~Nitta, K.~Ohashi and D.~Tong,
  ``Skyrmions from instantons inside domain walls,''
  Phys.\ Rev.\ Lett.\  {\bf 95}, 252003 (2005)
  [hep-th/0508130];
  M.~Eto, T.~Fujimori, M.~Nitta, K.~Ohashi and N.~Sakai,
  ``Domain walls with non-Abelian clouds,''
  Phys.\ Rev.\ D {\bf 77}, 125008 (2008)
  [arXiv:0802.3135 [hep-th]].


\bibitem{Nitta:2015mma} 
  M.~Nitta,
  ``Josephson junction of non-Abelian superconductors and non-Abelian Josephson vortices,''
  arXiv:1502.02525 [hep-th].

\bibitem{Nitta:2015mxa} 
  M.~Nitta,
  ``Josephson instantons and Josephson monopoles in a non-Abelian Josephson junction,''
  arXiv:1503.02060 [hep-th].


\bibitem{Eto:2008yi} 
  M.~Eto, T.~Fujimori, S.~B.~Gudnason, K.~Konishi, M.~Nitta, K.~Ohashi and W.~Vinci,
  ``Constructing Non-Abelian Vortices with Arbitrary Gauge Groups,''  Phys.\ Lett.\ B {\bf 669}, 98 (2008)  [arXiv:0802.1020 [hep-th]].  

\bibitem{Eto:2009bg} 
  M.~Eto, T.~Fujimori, S.~B.~Gudnason, K.~Konishi, T.~Nagashima, M.~Nitta, K.~Ohashi and W.~Vinci,
  ``Non-Abelian Vortices in SO(N) and USp(N) Gauge Theories,''
  JHEP {\bf 0906}, 004 (2009)
  [arXiv:0903.4471 [hep-th]];
  L.~Ferretti, S.~B.~Gudnason and K.~Konishi,
  ``Non-Abelian vortices and monopoles in SO(N) theories,''
  Nucl.\ Phys.\ B {\bf 789}, 84 (2008)
  [arXiv:0706.3854 [hep-th]];
  M.~Eto, T.~Fujimori, S.~B.~Gudnason, M.~Nitta and K.~Ohashi,
  ``SO and US(p) Kahler and Hyper-Kahler Quotients and Lumps,''
  Nucl.\ Phys.\ B {\bf 815}, 495 (2009)
  [arXiv:0809.2014 [hep-th]];
  M.~Eto, T.~Fujimori, S.~B.~Gudnason, Y.~Jiang, K.~Konishi, M.~Nitta and K.~Ohashi,
  ``Vortices and Monopoles in Mass-deformed SO and USp Gauge Theories,''
  JHEP {\bf 1112}, 017 (2011)
  [arXiv:1108.6124 [hep-th]].



\end{thebibliography}
\end{document}